\documentstyle[twoside,fleqn,npb,epsfig]{article}
\newcommand{\PO}{I\!\!P}

\newcommand{\xpom}{x_{\PO}}
\newcommand{\AmS}{{\protect\the\textfont2
  A\kern-.1667em\lower.5ex\hbox{M}\kern-.125emS}}

\title{Measurement  
of the diffractive structure function $F_2^{D(3)}$ 
}
\author{Christophe Royon (for the H1 Collaboration) 
\address{Service de Physique des Particules, DAPNIA, CEA-Saclay \\
91191 Gif sur Yvette Cedex, France%
        }}

\begin{document}

\begin{abstract}
Measurements of the diffractive structure function 
$F_2^{D(3)}(x_{\PO}, \beta, Q^2)$, 
describing the process $ep \rightarrow eXY$,
are presented in the two kinematic regions $0.4 \leq Q^2 \leq 5$,
${\rm GeV}^2$, $0.001 \leq \beta \leq 0.65$, and 
$200 \leq \beta \leq 800 \ {\rm GeV^2}$, $0.4 \leq \beta \leq 0.9$, both
with $\xpom < 0.05$, 
$M_{_Y} < 1.6 \ {\rm GeV}$ and $|t| < 1.0 \ {\rm GeV^2}$.
Together with published measurements at intermediate $Q^2$, the data are
compared with models based on QCD and Regge phenomenology.
The diffractive trajectory is found to have an intercept
larger than that describing soft hadronic data and 
consistent with that determined using previously
published H1 measurements at intermediate $Q^2$ alone.
The data are also parameterised 
using a QCD motivated model
based on the exchange of two gluons from the proton. In this model, the higher
twist contribution to $F_2^{D(3)}$ at large $\beta$ is found to be important 
at low $Q^2$. The data are also compared with models based on BFKL
dynamics.
\end{abstract}

\maketitle

\section{Diffractive structure function measurement}
In this paper, we report measurements of the diffractive structure function 
$F_2^{D(3)} (\beta, Q^2, \xpom)$ based on deep-inelastic scattering (DIS) data 
taken between 1995 and 1997 by the H1 collaboration.
The $F_2^{D(3)}$ measurement describes the process $ep \rightarrow eXY$.
As in our previous analysis of 1994 data \cite{F2d94}, the
two distinct hadronic systems $X$ and $Y$ are separated by the largest 
interval in rapidity between final state hadrons. 
The system $Y$ is closest to the direction of the outgoing proton beam.

The kinematics of the process can be described by
the invariant masses $M_X$ and $M_Y$
of the systems $X$ and $Y$, and the Lorentz scalars
\begin{eqnarray}
&~& x=\frac{-q^2}{2P \cdot q} \hspace{1.5cm} y=\frac{P \cdot q}{P \cdot k}
\\ 
&~& Q^2=-q^2 \hspace{1.5cm} t=(P-P_Y)^2 \ ,
\end{eqnarray}
where $P$ and $k$ are the 4-momenta of the incident proton and electron respectively,
$P_Y$ is the 4-momentum of subsystem $Y$ and $q$ is the 4-momentum of the
exchanged virtual photon coupling to the electron. The measurements presented
here are corrected to the region $M_Y < 1.6 \ {\rm GeV}$ and 
$|t| < 1.0 \ {\rm GeV^2}$. The 
following variables are also defined:
\begin{eqnarray}
&~& \beta= \frac{-q^2}{2q \cdot (P-P_Y)} = \frac{Q^2}{Q^2+M_X^2-t} 
\\
&~& x_{\PO}= \frac{q \cdot (P-P_Y)}{q \cdot P} 
=\frac{x}{\beta} \ ,
\label{xpombeta}
\end{eqnarray}
where $W^2=(q+P)^2$ is the center of mass energy squared of the virtual 
photon-proton system, $M_P$ the proton mass and $x$ is the Bjorken scaling 
variable.

During the 1994/95 HERA shutdown, 
the backward region of the H1 detector \cite{h1detector}
(the direction of the outgoing electron beam) was upgraded,
allowing an extension of the measurement to lower $Q^2$ 
and $\beta$ values compared to 1994 \cite{tim}.
Using data taken during a period when the interaction
vertex was shifted by $70 \ {\rm cm}$ in the proton beam direction,
the accessible kinematic range is extended still further. The minimum value 
of $Q^2$ at which $F_2^D$ is measured is thus reduced
by a factor 10 and the minimum
$\beta$ by a factor 40 compared to the previous measurement
\cite{F2d94}. 

The good running of the HERA machine in recent years has enabled a
large increase in the integrated luminosity recorded by H1. One impact
of this increase in statistics is the extension of the 
accessible kinematic range to higher values of
$Q^2$. A measurement of $F_2^{D(3)}$ for $200\le Q^2\le 800$~${\rm GeV^2}$
is presented in this paper, based on positron-proton
scattering data collected in the years 1995-1997. This data sample
represents an increase in statistics by a factor of around 15 compared to 
previous H1 $F_2^{D(3)}$ measurements~\cite{F2d94}.

The structure function $F_2^{D(3)}$
is calculated from the differential cross section according to the formula
\begin{eqnarray}
F_2^{D(3)}= 
\frac{\beta^2 Q^4}{4 \pi \alpha_{em}^2} \
\frac{1}{(1-y-\frac{y^2}{2})} \ \frac{{\rm d}^3 \sigma_{ep \rightarrow 
eXY}}{{\rm d}Q^2 \, {\rm d} \beta \, {\rm d} x} \ ,
\end{eqnarray}
taking the ratio $R$ of the longitudinal to the transverse cross sections to
be 0. 

\begin{figure}[p]
\begin{center}
\psfig{figure=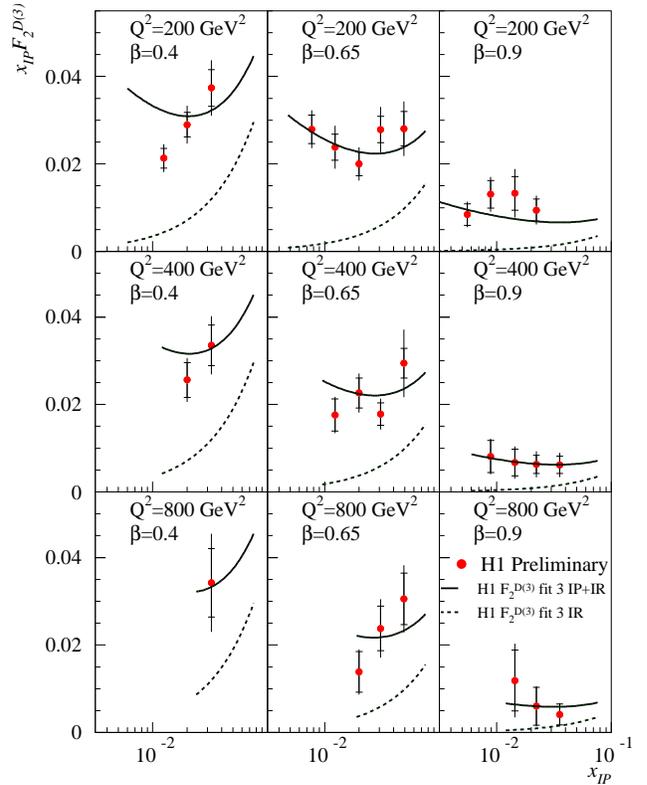 ,height=4.5in}
\end{center}
\caption{$x_{\PO} \cdot F_2^{D(3)}$ for the large $Q^2$ measurement, shown as 
a function of $\xpom$ in bins of $Q^2$ and $\beta$. The inner errors bars
show statistical errors only. The outer error bars show the statistical
and systematic  uncertainties added in quadrature.
The data are compared to a QCD fit to intermediate $Q^2$ data with parton
distributions for the pomeron and sub-leading exchange evolved into the
large $Q^2$ region using the DGLAP equations.}
\label{hiq2h1}
\end{figure}

\begin{figure}[p]
\begin{center}
\psfig{figure=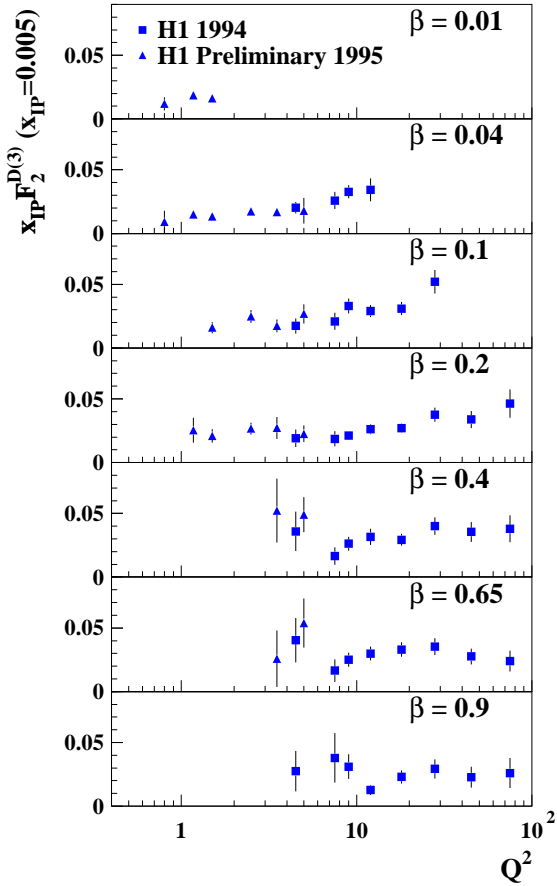 ,height=5in}
\end{center}
\caption{The structure function
$x_{\PO} \cdot F_2^{D(3)}$ ($x_{\PO}=0.005$), presented 
as a function of $Q^2$ in
bins of $\beta$, over the full $Q^2$ range accessed
with the 1994 and low $Q^2$ 1995 data sets. 
}
\label{qcdplot}
\end{figure}

The high $Q^2$ data are shown in the form of $x_{\PO} \cdot F_2^{D(3)}$ 
in figure~\ref{hiq2h1}.
Due to the kinematic constraints imposed by the large $Q^2$ values, the
measurements are restricted to $\beta \geq 0.4$ and relatively large
$\xpom$. In Figure~\ref{qcdplot}, the  
$Q^2$ and $\beta$ dependences of the 1994 and low $Q^2$ 1995
measurements of \linebreak $F_2^{D(3)}(Q^2,\beta,x_{\PO})$
are shown at a fixed small value of $x_{\PO}$ ($x_{\PO}=0.005$), for
which sub-leading exchange 
contributions are small. We clearly see changes of slopes
of scaling violations at low $Q^2$ ($Q^2<3$ GeV$^2$). In the large $\beta$,
low $Q^2$ region, additional structures are observed. These are presumably
related to higher twist contributions, in particular, the resonant production
of vector mesons.

\section{Interpretation and comparison with models}
The results are first used to further constrain the Regge phenomenological 
model investigated previously \cite{F2d94}.  A fit to the combined 
1994/1995 points of a single pomeron trajectory with intercept 
$\alpha_{I\!\!P}(0)$ is not able to describe the data.  Addition of a 
sub-leading (reggeon) trajectory with independent normalisation yields 
a significantly better description. 
The intercept trajectories are 
consistent with the published values \cite{F2d94}.  Fits with zero 
and maximal interference between the two trajectories describe the 
data equally well.  An extension to the fit in which the pomeron 
intercept takes the form $\alpha_{I\!\!P}(0) = a + b \log Q^2$ gives 
a value of $b$ compatible with zero.

The scaling violations of $F_2^{D(3)}$ as a function of $Q^2$ 
motivate an analysis of the data in which the $Q^2$ and 
$\beta$ dependence of the structure function is understood in terms of parton 
distribution functions for the pomeron, evolved with perturbative QCD. 
The quark flavour-singlet and gluon distributions are evolved in $Q^2$ 
with the NLO DGLAP equations and fitted, 
in combination with a reggeon contribution, to the combined data.  
The extracted parton density functions indicate a large gluonic 
content ($80-90\%$) of the pomeron.

\begin{figure}[p]
\begin{center}
\psfig{figure=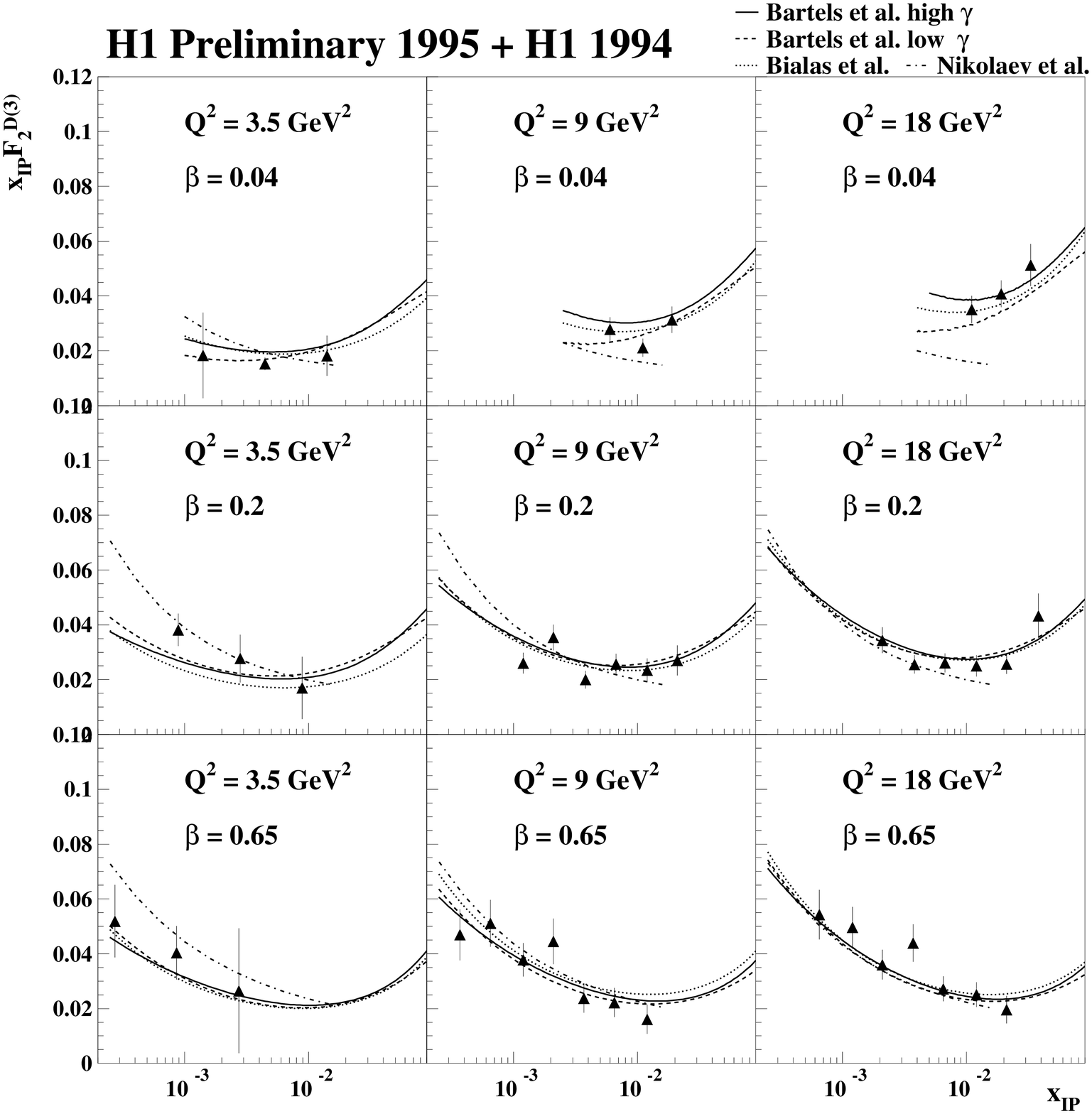 ,height=3.7in}
\end{center}
\caption{A subset of 1994 and low $Q^2$ 
1995 measurements of the structure function 
$x_{\PO} F_2^{D(3)}$, showing the $\xpom$ dependence in bins of $Q^2$ and 
$\beta$. The statistical errors are shown added in quadrature
with those systematic errors that vary between data points. Overall
normalisation uncertainties of 4.7\% for the 1995 data and 6.0\% for the 
1994 data are not shown. The data
are compared to the results of a fit to the Bartels et al. two-gluon
exchange model (high $\gamma$ solution: full line, low $\gamma$ solution:
dashed line), to the fit based
on the dipole model of Bialas et al. (dotted line) and to the prediction
of the Nikolaev et al. model (dashed-dotted line). 
For all models except that of Nikolaev et al.,
an additional sub-leading trajectory is added as described in the text. 
}
\label{bart}
\end{figure}

Then we consider the two gluon exchange model for interaction.
In a recent paper \cite{bartels}, a parameterization of the diffractive 
structure function in 
terms of three main contributions was proposed. The photon
fluctuates into partonic states which scatter diffractively.
At the beginning of the scattering process, the photon splits into
a $q \bar{q}$ pair, and at sufficient $M_X^2$, the $q \bar{q}$ pair can 
radiate an additional gluon before it reaches the proton at rest. At small
diffractive masses, it is expected that the longitudinal cross section for
$q \bar{q}$ pair production is not small compared to the transverse cross
section, this third term appear to be a higher twist 
contribution.
A fit is performed to the combined 1994 and 1995 data with the
restriction $Q^2 > 3$ $\rm{GeV}^2$ to remain in the domain
of perturbative QCD. Two solutions are found (see figure \ref{bart})
corresponding to large and low $\gamma$ where $\gamma$ is a parameter which
describes the $\beta$ dependance of the 
$q \bar{q} g$ contribution.

The 1994 and 1995 measurements of  
$F_2^{D(3)}$ are also compared with two additional models, 
the QCD dipole model in which the diffractive
interaction is based on BFKL \cite{bfkl} dynamics
 \cite{pesh} and the model of Nikolaev et al.  
\cite{Nikolaev} (see figure \ref{bart}). The distinction between these models
will be hardly feasible using $F_2^D$ data alone. Final state and longitudinal
structure function measurement may give more hints to distinguish between them.

\end{document}